\documentclass[aps,prl,twocolumn,floatfix]{revtex4}
\usepackage{graphicx}
\usepackage{epstopdf}
\usepackage{bm}

\usepackage[latin1]{inputenc}
\usepackage{fontenc}
\usepackage{amsmath}
\usepackage{amssymb}

\usepackage{shadow}
\usepackage{pstricks}

\def\Tr{\mathop{\rm Tr}\nolimits}

\newcommand {\be}[1]{\begin{eqnarray} \mbox{$\label{#1}$}  }
\newcommand{\ee}{\end{eqnarray}}

\newcommand{\pref}[1]{(\ref{#1})}

\newcommand{\bs}{\boldsymbol}

\newcommand{\ket}[1]{|#1\rangle}
\newcommand{\bra}[1]{\langle #1 |}

\newcommand{\cD}{ {\cal D} }
\newcommand{\cP}{ {\cal P} }
\newcommand{\cS}{ {\cal S} }

\begin{document}
\title{Extreme points of the set of density matrices with positive
  partial transpose}
\author{J.~M.~Leinaas}
\affiliation{Department of Physics, University of Oslo, N-0316 Oslo,
  Norway}
\author{J.~Myrheim}
\affiliation{Department of Physics, 
The Norwegian University of Science and Technology, 7491 Trondheim, Norway}
\author{E.~Ovrum}
\affiliation{Department of Physics and Center of Mathematics for
  Applications, 
University of Oslo, N-0316 Oslo, Norway}
\date{April 24, 2007}
\begin{abstract}
We present a necessary and sufficient condition for a finite
dimensional density matrix to be an extreme point of the convex set of
density matrices with positive partial transpose with respect to a
subsystem. We also give an algorithm for finding such extreme points
and illustrate this by some examples.

\end{abstract}

\maketitle


The density matrices describing states of a bipartite quantum system
are uniquely classified as being either {\em separable} or {\em
entangled}.  However, it may be a non-trivial task in practice to
classify a given density matrix.  One simple method which can
sometimes prove a density matrix to be entangled, and which is
especially useful in systems with Hilbert spaces of low dimensions, is
due to A.~Peres \cite{PeresCriterion}.  Peres noted that a separable
density matrix must remain positive when a partial transposition is
performed with respect to one of the subsystems.  In general the set
of density matrices that have positive partial transpose, called PPT
states for short, is larger than the set of separable matrices, but in
low dimensions the difference between these two sets is small.  In
particular, for a system composed of one subsystem of dimension 2 and
one subsystem of dimension 2 or 3, the two sets are known to be
identical \cite{Horodecki96,Horodecki97}.

A particular approach to the question of separability is to
characterize geometrically the different sets of matrices as subsets
of the real Hilbert space of hermitian matrices~\cite{bengtsson}.
Thus, the set of all density matrices is a compact convex set $\cD$,
with the set of separable density matrices, $\cS$, and the set of PPT
matrices, $\cP$, which we will call the {\em Peres set}, as convex
subsets.  The Peres criterion states that $\cS\subset\cP\subset\cD$.
We have presented elsewhere a numerical method for deciding whether 
a density matrix is included in $\cS$ by computing the distance to the closest
separable density matrix~\cite{lmo,dlmo}.

There are two complementary ways to characterize a convex set.  One
way is to specify its {\em extreme points}.  Thus, every point in a
compact convex set of finite dimension 
$d$ 
has an expansion as a
convex combination of 
$d+1$ 
extreme points, while an extreme point
cannot be written as a convex combination of other points in the set.
The other way is to identify the set by conditions, typically
algebraic equations and inequalities, that define whether or not a
given point belongs to the set.  For example, the convex set $\cD$ may
be defined by its extreme points, which are the pure states (one
dimensional projections), or equivalently by the conditions of
hermiticity, positivity and unit trace.  In general there is no simple
relation between these two descriptions of a convex set.

It is interesting to note that the convex sets $\cS$ and $\cP$ have
simple charcterizations in the complementary ways just described.
Thus, the extreme points of $\cS$ are well known, they are the pure
product states of the two subsystems, but there is no (known)
effective test for membership of $\cS$.  In contrast, testing for
membership of $\cP$ is easy, because $\cP=\cD\cap\cD^P$, where the
superscript $P$ denotes partial transposition, but the extreme points
of $\cP$ are not completely known.

The purpose of the present paper is to study the difference between
the two sets $\cS$ and $\cP$ by studying the extreme points of $\cP$.
The extreme points of $\cS$ are also extreme points of $\cP$, because
they are extreme points of $\cD$ and $\cS\subset\cP\subset\cD$.  But
in general $\cP$ is larger than $\cS$ and has additional extreme
points, which are unknown and which make the whole difference between
the two sets.

The extreme points of $\cP$ that are not pure product states are
interesting also as examples of entangled PPT states.
%
%
In fact, because $\cS\subset\cP$, a separable extreme point of $\cP$
must be an extreme point of $\cS$ and hence a pure product state.  It
is easy to verify that any pure state which is not a product state, is
not in $\cP$ and is therefore entangled.  Thus, every extreme point of
$\cP$ which is not a pure product state, is entangled and has rank
higher than one.  A stronger lower bound on the ranks of entangled PPT
states is actually known \cite{minRank}.

We will in the following specify a criterion for uniquely identifying
the extreme points of $\cP$ and we will describe an algorithm for
finding such points. The method is demonstrated by some examples.

To be more specific we consider a composite quantum system with
Hilbert space
${\cal H} = {\cal H}_A \otimes {\cal H}_B$ of finite dimension
$N=N_A\,N_B$, where $A$ and $B$ denote the two subsystems.  Then $\cD$
is the set of density matrices on ${\cal H}$,
\be{l1}
\rho \in {\cal D} \Leftrightarrow  
\rho=\rho^\dag, \quad
\rho \geq 0, \quad
\Tr\rho=1
\ee
A subset is the Peres set $\cP$,
\be{l2}
\rho \in {\cal P} \Leftrightarrow
\rho \in {\cal D}\; \text{and}\;
\rho^P \in {\cal D}
\ee
where $\rho^P$ is the partial transpose of $\rho$ with respect to any
one of the subsystems, say system $B$. A subset of $\cP$ is $\cS$, the
set of separable density matrices,
\be{l3}
\rho \in {\cal S} \Leftrightarrow  \rho=\sum_k p_k \rho_k^A\otimes\rho_k^B,
\quad p_k>0,
\quad \sum_k p_k=1
\ee
Thus, $\cS$ is the {\em convex hull} of the product states of the two
subsystems, and its extreme points are the {\em pure} product states.

To develop the method we need for finding the extreme points of $\cP$
we will first apply it to $\cD$, the full set of density matrices, with
the pure states as extreme points.  We recall some definitions and
elementary facts.

Every $\rho\in\cD$ is hermitian and has a spectral decomposition
\be{l4}
\rho=\sum_i \lambda_i\,\ket{\psi_i}\bra{\psi_i}
\ee
with real eigenvalues $\lambda_i$ and orthonormal eigenvectors
$\ket{\psi_i}$.  The positivity condition $\rho\geq 0$ means that all
$\lambda_i\geq 0$, or equivalently that
$\bra{\psi}\rho\ket{\psi}\geq 0$ for all $\ket{\psi}$, with
$\bra{\psi}\rho\ket{\psi}=0$ if and only if $\rho\ket{\psi}=0$.  The
orthogonal projection
\be{l5}
P=\sum_{i,\lambda_i>0} \ket{\psi_i}\bra{\psi_i}
\ee
projects onto the {\em image} (or {\em range}) of $\rho$, whereas
$1-P$ projects onto the {\em kernel} of $\rho$, denoted by $\ker\rho$.

If $\rho$ is not an extreme point of $\cD$ it is a convex combination
\be{l6}
\rho = x\rho'+ (1-x)\rho''\,,\quad 0<x<1
\ee
with $\rho',\rho''\in{\cal D}$ and $\rho'\neq\rho''$.  The identity
\be{l7}
\bra{\psi}\rho\ket{\psi} =
x\,\bra{\psi}\rho'\ket{\psi}+ (1-x)\,\bra{\psi}\rho''\ket{\psi}
\ee
shows that $\rho\geq 0$ when $\rho'\geq 0$ and $\rho''\geq 0$, thus
proving the convexity of $\cD$.  More interestingly, it shows that
\be{l8}
\ker\rho=\ker\rho'\cap\ker\rho''
\ee
%
With $P$ defined as in~\pref{l5} we have therefore $P\rho P=\rho$,
$P\rho' P=\rho'$, and $P\rho'' P=\rho''$.

When~\pref{l6} holds, the matrix $\sigma=\rho'-\rho$ is hermitian and
nonzero, and $\Tr\sigma=0$, hence $\sigma$ has both positive and
negative eigenvalues.  Moreover, $P\sigma P=\sigma$.
Define
\be{l9}
\tau(x)=\rho+x\sigma
\ee
for $x$ real.  Since $\sigma$ has both positive and negative
eigenvalues, so has $\tau(x)$ for large enough $|x|$.  If
$\rho\ket{\psi}=0$ then $P\ket{\psi}=0$ and
$\sigma\ket{\psi}=P\sigma P\ket{\psi}=0$, hence $\tau(x)\ket{\psi}=0$
for all $x$.  Therefore only the strictly positive eigenvalues of
$\rho$ can change when $x\sigma$ is added to $\rho$, and since they
change continuously with $x$, they remain positive for $x$ in a finite
interval about $x=0$.

We conclude that there exists an $x_1<0$ and an $x_2>0$ such that
$\tau(x)\geq 0$ for $x_1\leq x\leq x_2$, and $\tau(x)\not\geq 0$ for
$x<x_1$ or $x>x_2$.  At $x=x_1$ or $x=x_2$, $\tau(x)$ has at least
one zero eigenvalue more than $\rho$.

We are now prepared to search systematically for extreme points of
$\cD$.  Starting with an arbitrary $\rho_1\in\cD$ we define the
projection $P_1$ in the same way as we defined $P$ from $\rho$
in~\pref{l5}.  If we can find a hermitian matrix $\sigma$ solving the
equation
\be{l10}
P_1\sigma P_1=\sigma
\ee
we define
\be{l11}
\sigma_1=\sigma-(\Tr\sigma)\,\rho_1
\ee
in order to have $P_1\sigma_1P_1=\sigma_1$ and $\Tr\sigma_1=0$.
Clearly $\sigma=\rho_1$ is a solution of~\pref{l10}.  If this is the
only solution,
then only $\sigma_1=0$ is possible, and it follows from the above
discussion that $\rho_1$ is an extreme point.  The number of linearly
independent solutions of~\pref{l10} is $n_1^{\;2}$, where $n_1=\Tr
P_1$ is the rank of $\rho_1$.  Hence, $\rho_1$ is an extreme point if
and only if $n_1=1$ so that it is a pure state.

If $\rho_1$ is not an extreme point, then we can find $\sigma_1\neq 0$
and define
\be{l12}
\tau_1(x)=\rho_1+x\sigma_1
\ee
We increase (or decrease) $x$ from $x=0$ until it first happens that
$\tau_1(x)$ gets one or more additional zero eigenvalues as compared
to $\rho_1$.  We will know if we go too far, because then $\tau_1(x)$
will get negative eigenvalues.  We choose $\rho_2$ as the limiting
$\tau_1(x)$ determined in this way.  By construction, $\rho_2$ has
lower rank than $\rho_1$.

We repeat the whole procedure with $\rho_2$ in place of $\rho_1$, and
if $\rho_2$ is not extreme we will find a $\rho_3$ of lower rank.
Continuing in the same way, we must end up at an extreme point
$\rho_K$, with $K\leq N$, since we get a decreasing sequence of
projections,
\be{l13}
I\supseteq P_1\supset P_{2}\supset \ldots \supset P_K
\ee
of decreasing ranks
%
$N\geq n_1>n_2>\ldots>n_K=1$.

We may understand the above construction geometrically.  In fact, each
projection $P_k$ defines a convex subset $P_k{\cal D}P_k$ of $\cD$,
with $\rho_k$ as an interior point.  This subset consists of all
density matrices on the $n_k$ dimensional Hilbert space $P_k{\cal H}$,
and if $n_k<N$ it is a flat face of the boundary of $\cD$.


It is straightforward to adapt the above method and use it to search
for extreme points of $\cP$.  Thus, we consider an initial matrix
$\rho_1\in\cP$, characterized by two integers $(n_1,m_1)$, the ranks
of $\rho_1$ and of the partial transpose $\rho_1^P$, since
$\rho_1\in\cP$ means that $\rho_1\in\cD$ and $\rho^P_1\in\cD$.  We
denote by $P_1$ the projection on the image of $\rho_1$, as before,
and we introduce $Q_{1}$ as the projection on the image of
$\rho^P_{1}$.  We have to solve the two equations
\be{PQcond}
P_1\sigma P_1=\sigma\,,\quad Q_1\sigma^P Q_1=\sigma^P
\ee
To understand these equations it may help to think of $\sigma$ as a
vector in the $N^2$ dimensional real Hilbert space ${\cal M}$ of
$N\times N$ hermitian matrices with the scalar product
\be{l16}
\langle A,B\rangle=\langle B,A\rangle=
\Tr(AB)=\sum_{i,j}A^{\;\ast}_{ij}\,B_{ij}
\ee
If ${\bf L}$ is a linear transformation on ${\cal M}$, its transpose
${\bf L}^T$ is defined by the identity
%
$\langle A,{\bf L}^TB\rangle=\langle {\bf L}A,B\rangle$,
%
and ${\bf L}$ is symmetric if ${\bf L}^T={\bf L}$.  Partial
transposition of $A\in{\cal M}$ permutes the matrix elements of $A$
and is a linear transformation
%
$\bs\Pi A=A^P$.
%
It is its own inverse,
and is an orthogonal transformation (it preserves the scalar product),
%
%
hence $\bs\Pi=\bs\Pi^{-1}=\bs\Pi^T$.  It also preserves the trace,
$\Tr(\bs\Pi A)=\Tr A$.  The projections $P_ {1}$ and $Q_{1}$ on ${\cal
H}$ define projections ${\bf P}_{1}$ and ${\bf Q}_{1}$ on ${\cal M}$
by
%
\be{l19}
{\bf P}_1A=P_1AP_1\,,\quad
{\bf Q}_1A=Q_1AQ_1
\ee
These are both orthogonal projections:
${\bf P}_{1}^{\;2}={\bf P}_{1}$,
${\bf P}_{1}^T={\bf P}_{1}$,
${\bf Q}_{1}^{\;2}={\bf Q}_{1}$, and
${\bf Q}_{1}^T={\bf Q}_{1}$
%

In this language the equations~\pref{PQcond} may be written as
%
%
%
\be{PQcond2}
{\bf P}_{1}\sigma=\sigma\,,\quad {\bf \bar Q}_{1}\sigma=\sigma
\ee
with ${\bf \bar Q}_{1}=\bs\Pi {\bf Q} _{1}\bs\Pi $.
Note that ${\bf\bar Q}_1$ is also an orthogonal projection:
${\bf\bar Q}_1^{\;2}={\bf\bar Q}_1$ and
${\bf\bar Q}_1^T={\bf\bar Q}_1$.
These two equations
are equivalent to the single equation
\be{l22}
{\bf P}_{1}{\bf \bar Q}_{1}{\bf P}_{1}\sigma
=\sigma
\ee
or equivalently
${\bf \bar Q}_{1}{\bf P}_{1}{\bf \bar Q}_{1}\sigma=\sigma$.  They
restrict the hermitian matrix $\sigma$ to the intersection between
the two subspaces of ${\cal M}$ defined by the projections
${\bf P}_{1}$ and ${\bf \bar Q}_{1}$.  We shall denote by
${\bf B}_{1}$ the projection on this subspace, which is spanned by the
eigenvectors with eigenvalue $1$ of
${\bf P}_{1}{\bf \bar Q}_{1}{\bf P}_{1}$.  Because the latter is a
symmetric linear transformation on ${\cal M}$ it has a complete set of
orthonormal real eigenvectors and eigenvalues.  All its eigenvalues
lie between 0 and 1.  We may diagonalize it in order to find its
eigenvectors with eigenvalue $1$.

Having found $\sigma$ as a solution of~\pref{l22} we define $\sigma_1$
as in~\pref{l11}.  If $\sigma=\rho_1$ and $\sigma_1=0$ is the only
solution, then $\rho_1$ is an extreme point of $\cP$.  If we can find
$\sigma_1\neq 0$, then we define $\tau_1(x)$ as in~\pref{l12}, and
increase (or decrease) $x$ from $x=0$ until we reach the first value
of $x$ where either $\tau_1(x)$ or $(\tau_1(x))^P$ has at least one
new zero eigenvalue.  This special $\tau_1(x)$ we take as $\rho_2$.
By construction, when $n_2$ is the rank of $\rho_2$ and $m_2$ the rank
of $\rho_2^P$, we have $n_2\leq n_1$, $m_2\leq m_1$, and either
$n_2<n_1$ or $m_2<m_1$.

Next, we check whether $\rho_2$ is an extreme point of $\cP$, in the
same way as with $\rho_1$.  If $\rho_2$ is not extreme, then we can
find a third candidat $\rho_3$, and so on.  We will reach an extreme
point $\rho_K$ in a finite number of iterations, because the sum of
ranks, $n_k+m_k$, decreases in each iteration.

The geometrical interpretation of this iteration scheme is that the
density matrices $\rho\in\cP$ which satisfy the condition
${\bf B}_k\,\rho=\rho$ define a convex subset of $\cP$ having $\rho_k$
as an interior point.  This subset is either the whole of $\cP$, or a
flat face of the boundary of $\cP$, which is the intersection of a
flat face of $\cD$ and a flat face of $\cD^P$.

The construction discussed above defines an algorithm for finding
extreme points of $\cP$, and gives at the same time a necessary and
sufficient condition for a density matrix in $\cP$ to be an extreme
point.  Let us restate this condition:

{\em A density matrix $\rho\in\cP$ is an extreme point of $\cP$ if and
  only if the projection $P$ which projects on the image of $\rho$ and
  the projection $Q$ which projects on the image of $ \rho^P$, define
  a combined projection $\bf B$ of rank 1 in the real Hilbert space
  ${\cal M}$ of hermitian matrices.}

The algorithm deserves some further comments. In iteration $k$ the
projections $P_k$, $Q_k$ and ${\bf B}_k$ are uniquely defined by the
density matrix $\rho_k$, but the matrix $\sigma_k$
is usually not unique. In the applications discussed below we
have simply chosen $\sigma_k$ randomly.  Clearly, more systematic
choices are possible if one wants to search for special types of
extreme points.

Another comment concerns the ranks $(n,m)$ of a density matrix $\rho$
and its partial transpose $\rho^P$ if it is an extreme point.  We can
derive an upper limit on these ranks.  The rank of the projection
${\bf P}$ is $n^2$ and the rank of ${\bf \bar Q}$ is $m^2$, thus, the
equations ${\bf P}\sigma=\sigma$ and ${\bf\bar Q}\sigma=\sigma$ for
the $N^2$ dimensional vector $\sigma$ represent $N^2-n^2$ and
$N^2-m^2$ constraints, respectively.  The total number of {\em
independent} constraints is $n_c\leq 2N^2-n^2-m^2$, and the rank of
${\bf B}$ is $N^2-n_c\geq n^2+m^2-N^2$.  For an extreme point the rank
of ${\bf B}$ is $1$, implying the inequality
\be{constraint}
n^2+m^2\leq N^2+1
\ee

We have used our algorithm in numerical studies.  In one approach
we use as initial density matrix the maximally mixed state
$\rho_1=1/N$ and choose in iteration $k$ a random direction
in the subspace ${\bf B}_k{\cal M}$.  We list in Table~1 all the ranks
of extreme points found in this way in various dimensions
$N=N_AN_B$.  We do not distinguish between ranks $(n,m)$ and $(m,n)$
since there is full symmetry between $\rho$ and $\rho^P$.

\begin{table}[h!]
\begin{center}
 \begin{tabular}{c|cc|c}\hline\hline
$N_A\times N_B=N$ &\multicolumn{2}{c|}{$(n,m)$}   & $n+m$  \\ \hline
$2\times 4=\phantom{1}8$ & (5,6) & & 11 \\
$3\times 3=\phantom{1}9$ &  (6,6) & (5,7) & 12\\
$2\times 5=10$ & (7,7) & (6,8) & 14\\
$2\times 6=12$ & (8,9) & & 17\\
$3\times 4=12$ & (8,9) & & 17\\
$3\times 5=15$ & (10,11) & & 21\\
$4\times 4=16$ &\,(11,11)\,&\,(10,12)\,& 22 \\
$3\times 6=18$ & (12,13) & & 25\\
$4\times 5=20$ & (14,14) & (13,15) & 28\\
$5\times 5=25$ & (17,18) & & 35\\
\end{tabular}
\end{center}
\caption{Typical ranks $(n,m)$ for extreme points}
\label{fig:extremeRank}
\end{table}

We find only solutions of {\em maximal rank}, in the sense that
increasing either $n$ or $m$ will violate the
inequality~\pref{constraint}.  Maximal rank means that the constraints
given by the equations ${\bf P}\sigma=\sigma$ and
${\bf\bar Q}\sigma=\sigma$ are mostly independent.  Furthermore, we
find only the most symmetric ranks, in the sense that $m\approx n$.
For example, in the $4\times 4$ system we find ranks $(11,11)$ and
$(10,12)$, but not $(9,13)$, $(7,14)$ or $(5,15)$, which are also
maximal.

\begin{figure}[h]
\label{veier}
\begin{center}
\includegraphics[height=5cm]{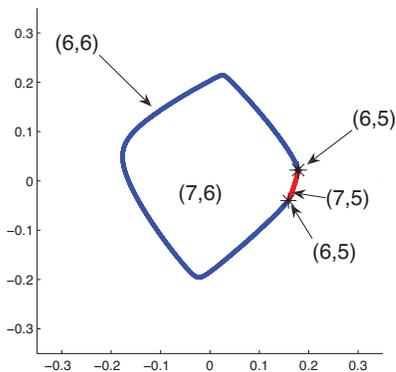}
\end{center}
\caption{\small Section through the boundary of $\cP$, in $3\times 3$
  dimensions.  A closed curve of extreme points surrounds a region of
  entangled PPT matrices of rank $(7,6)$.  The curve has two parts,
  characterized by ranks $(7,5)$ and $(6,6)$, joining at two rank
  $(6,5)$ extreme points.}
\end{figure}

The $3\times3$ system we have examined further in the following way.
For one specific sequence $\rho_1,\rho_2,\ldots,\rho_K$ with $\rho_K$
extreme, we repeat the final step, keeping $\rho_ {K-1}$ fixed but
choosing different directions in the subspace
${\bf B}_{K-1}{\cal M}$.  We find that every direction points directly
towards an extreme point.  Thus, $\rho_{K-1}$ is an interior point of
a flat face of the boundary of $\cP$ bounded by a hypersurface of
extreme points.  Fig.~1 shows a two dimensional section through this
flat face.

This shows that extreme points of non-maximal rank do exist, and the
fact that we do not find them in random searches just indicates that
they define subsets of lower dimension than the extreme points of
maximal rank $(n,m)$ with $m\approx n$.  Note that an extreme point
which is not a pure product state cannot have arbitrarily low rank,
since in
%
%
\cite{minRank}
there is a proof that all PPT matrices of rank
less than $N_0\equiv\min\{N_A,N_B\}$ are separable.  This implies a
lower limit of $(N_0,N_0)$ for the ranks of an extreme point of $\cP$
which is not a pure product state.  It is not known whether there
exist entangled PPT states of the minimal rank $N_0$.


Several of the low rank entangled PPT states that are known turn out, in our test,
to be extreme points.  Examples in $3\times 3$ dimensions include the
unextendible product basis state of rank $(4,4)$ \cite{UPB}; another
state of rank $(4,4)$ \cite{Kill-chan}; and explicit examples of rank
$(5,5)$ and $(6,6)$ states~\cite{clarisse}.

The entangled PPT states first discovered~\cite{Horodecki97}, in $3\times 3$
dimensions to be specific, are not extreme points of $\cP$, but on the 
flat face defined by the corresponding projection {\bf B} they seem
to be completely surrounded by extreme points.  Figure~2 is a two
dimensional section chosen so as to show one such state (with
parameter value $a=0.42$, called here the ``Horodecki state''), as a
convex combination of two extreme points of $\cP$.  We would expect a
two dimensional section through two extreme points of $\cP$ to show
maximum difference between the sets $\cS$ and $\cP$.  Thus, this plot
illustrates the fact that the difference is indeed very small in
$3\times 3$ dimensions.

\begin{figure}[h]
\label{veier2}
\begin{center}
\includegraphics[height=5cm]{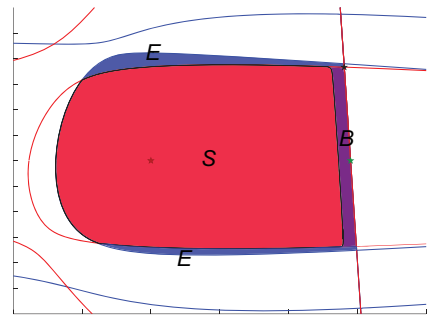}
\end{center}
\caption{\small Section through the set of density matrices in
  $3\times 3$ dimensions.  The star in the middle of a straight line
  is the ``Horodecki state'' (see text).  It is a convex combination
  of two extreme points, one of which is plotted as a star.  The
  maximally mixed state $1/N$ is the star close to the center.  The
  separable matrices lie in the large (red) region marked $S$, the
  entangled PPT states in the (purple) region marked $B$, and the
  entangled non-PPT states in the two (blue) regions marked $E$.  The
  lines are solutions of the equations $\det\rho=0$ (blue) and
  $\det\rho^P=0$ (red lines).}
\end{figure}


In conclusion, the method discussed has the potential of producing a
clearer picture of the difference between the two sets $\cS$ and $\cP$
and thereby the set of states with {\em bound entanglement}.  We
intend to follow up the work presented in this paper by other
numerical studies of composite systems of low Hilbert space
dimensions.

\begin{acknowledgments}
This work has been supported by NordForsk.
\end{acknowledgments}


\end{document}